\documentclass[11pt]{article}
\def \d {{\rm d}}
\def \e {{\rm e}}

\begin{document}

\title{Null limits of the C-metric}

\author{J. Podolsk\'y\thanks{E--mail: {\tt Podolsky@mbox.troja.mff.cuni.cz}}
\\
\\ Institute of Theoretical Physics, Charles University,\\
V Hole\v{s}ovi\v{c}k\'ach 2, 18000 Prague 8, Czech Republic.\\
\\
and J. B. Griffiths\thanks{E--mail: {\tt J.B.Griffiths@Lboro.ac.uk}} \\ \\
Department of Mathematical Sciences, Loughborough University \\
Loughborough, Leics. LE11 3TU, U.K. \\ }

\maketitle

\begin{abstract}
\noindent
The C-metric is usually understood as describing two black holes which
accelerate in opposite directions under the action of some conical
singularity. Here, we examine all the solutions of this type which
represent accelerating sources and investigate the null limit in which the
accelerations become unbounded. We show that the resulting space-times
represent spherical impulsive gravitational waves generated by snapping or
expanding cosmic strings. 


\end{abstract}

\section{Introduction}

The vacuum $C$-metric is a well-known solution of Einstein's equations. It
is described by the line element 
 \begin{equation}
 \d s^2 = -A^{-2}(x+y)^{-2} \left(F^{-1}\d y^2 + G^{-1}\d x^2 + G\d\phi^2 
-F\d t^2 \right),
 \label{Cmetric}
 \end{equation}
 where
 \begin{equation}
 F = -1 + y^2 - 2m Ay^3, \qquad G = 1 - x^2 - 2m Ax^3, 
 \label{FG}
 \end{equation}
 and $A, m$ are arbitrary constants. Kinnersley and Walker
\cite{KinWal70} (see also \cite{Bon83}) showed that this may represent two
black holes, each of mass $m$, that have uniform acceleration $A$ in
opposite directions. The acceleration is caused either by a strut between
the black holes or by two semi-infinite strings connecting them to
infinity.

The radiative properties of this space-time were investigated by Farhoosh
and Zimmerman \cite{FarZim80}, and its asymptotic properties by Ashtekar
and Dray \cite{AshDra81}. A transformation of the line element
(\ref{Cmetric}) into a form that explicitly exhibits its boost-rotation
symmetry \cite{BicSch89}, and which facilitates its physical
interpretation was achieved by Bonnor \cite{Bon83}. To maintain the
signature in (\ref{Cmetric}), $F$ and $G$ are required to be positive.
Assuming the condition that 
 \begin{equation}
 0\le |mA| <{1\over3\sqrt3}, 
 \label{rangemA}
 \end{equation}
the expressions in (\ref{FG}) have three real roots. This gives rise to
four possible space-times according to different ranges of the
coordinates $x$ and $y$. This has been discussed by Cornish and Uttley
\cite{CorUtt95}. However, in a recent review of these solutions, Pravda and
Pravdov\'a \cite{Pravdas00} have clarified that only three of these
represent space-times with accelerated sources.

The purpose of the present paper is to investigate the null limits of these
solutions as $A\to\infty$.

\section{The explicit solutions}

Particular solutions for the C-metric depend on the roots of the cubic 
 \begin{equation}
 2A^4z^3-A^2z^2+m^2=0,
 \label{cubic}
 \end{equation}
 which is related to the cubics in (\ref{FG}) by \ $z={m\over A}\,y$ \ or \
$z=-{m\over A}\,x$. The roots of (\ref{cubic}) are given by
 \begin{eqnarray}
 z_1 &=& {\textstyle{1\over6}\,A^{-2} \,\Big[\, 1+2\cos(\varphi+{2\over3}\pi)
\,\Big]}\>, \nonumber \\
 z_2 &=& {\textstyle{1\over6}\,A^{-2} \,\Big[\, 1+2\cos(\varphi+{4\over3}\pi)
\,\Big]}\>, \label{roots} \\
 z_3 &=& {\textstyle{1\over6}\,A^{-2} \,\Big[\, 1+2\cos\varphi \,\Big]}\>,
\nonumber 
 \end{eqnarray}
 where \ $\varphi={1\over3}\arccos(1-54m^2A^2)$. \ These roots satisfy \
$z_1\le z_2<z_3$ \ for all values of $mA$ in the permitted range
(\ref{rangemA}) which corresponds to \ $\varphi\in[\,0,{1\over3}\pi)$.

In coordinates adapted to the boost-rotation symmetry \cite{BicSch89}, the
line element (\ref{Cmetric}) takes the form 
 \begin{equation}
 \d s^2=-\e^\lambda \d\rho^2 - \rho^2\e^{-\mu} \d \phi^2 +
  (\zeta^2-\tau^2)^{-1} \left[\e^\mu (\zeta\d\tau-\tau\d\zeta)^2
-\e^\lambda(\zeta \d \zeta - \tau \d\tau)^2\right],
 \label{BSmetric}
 \end{equation}
 in which $\mu$ and $\lambda$ are specific functions of $\rho^2$ and
$\zeta^2-\tau^2$ (see \cite{Bon83}). For convenience of presenting
explicit solutions, let us define the following
expressions 
 \begin{eqnarray}
 && Z_1 = z_1-z_3 =-{\textstyle{1\over\sqrt3}\,A^{-2}
\sin(\varphi+{1\over3}\pi) }\>, \nonumber \\ 
 && Z_2 = z_2-z_3 =-{\textstyle{1\over\sqrt3}\,A^{-2}
\sin(\varphi+{2\over3}\pi) }\>, \nonumber \\
 && R ={\textstyle{1\over2}} (\zeta^2-\tau^2+\rho^2)\>, \nonumber \\
 && R_1 =\sqrt{(R+Z_1)^2-2Z_1\rho^2}\>, \qquad\quad\ 
  R_2 =\sqrt{(R+Z_2)^2-2Z_2\rho^2}\>,  \\
 && S_1 =R\,(R+Z_1+R_1)-Z_1\rho^2\>, \qquad
  S_2 =R\,(R+Z_2+R_2)-Z_2\rho^2\>, \nonumber \\
 && S_{12} =(R+Z_1)(R+Z_2)+R_1R_2-(Z_1+Z_2)\rho^2\>. \nonumber
 \end{eqnarray}
 Slightly modifying the notation of \cite{Pravdas00}, the three cases that
are of physical interest (denoted by $\cal{A}$, $\cal{B}$ and $\cal{D}$)
are now given by
 \begin{eqnarray}
 \e^\mu_{\cal{A}} = a\>
\frac{(R+Z_1+R_1-\rho^2)(R+Z_2+R_2-\rho^2)}{(\zeta^2-\tau^2)^2} \ , 
 &&\e^\lambda_{\cal{A}}= {a\over2}\, {S_1\,S_2\over R_1R_2\,S_{12}}\,,
\label{emuA}\\
 \e^\mu_{\cal{B}} = a\>
\frac{R+Z_1+R_1-\rho^2}{R+Z_2+R_2-\rho^2}  \ , \hskip102pt
 &&\e^\lambda_{\cal{B}}= {a\over2}\, {S_2\,S_{12}\over R_1R_2\,S_1}\,, 
\label{emuB}\\
 \e^\mu_{\cal{D}} = a\>
\frac{R+Z_2+R_2-\rho^2}{R+Z_1+R_1-\rho^2}  \ , \hskip102pt
 &&\e^\lambda_{\cal{D}}= {a\over2}\, {S_1\,S_{12}\over R_1R_2\,S_2}\,, 
\label{emuD}
 \end{eqnarray} 
 where $a$ is a positive constant. The condition that the metric is regular
on the ``roof'' $\zeta^2-\tau^2=0$ has already been inserted. The physical
interpretation of all these cases is described in \cite{Pravdas00}.

The case $\cal{A}$ describes two uniformly accelerated black holes with a
curvature singularity between them. In general, there is a conical
singularity on the axis $\rho=0$ extending from the black holes to
infinity. However, this is absent when $a=1$.

The physically most interesting case $\cal{B}$, which has been widely
considered in the literature, describes two uniformly accelerated black
holes connected to conical singularities. The axis is regular between the
particles when \ 
$a=Z_1/Z_2 =2\,\big[\,1+\sqrt3\cot(\varphi+{1\over3}\pi)\,\big]^{-1}$. \ In
this case, the black holes can be considered to be accelerated by two
strings connecting them to infinity. Alternatively, the axis is regular
outside the particles if $a=1$, in which case the black holes are
accelerated by a strut between them.

The case $\cal{D}$ also has similar conical singularities either between
the sources or connecting them to infinity, but the sources are now
different types of curvature singularities whose interpretation is
unclear. In this case, the axis is regular between the
singularities when \ $a=Z_2/Z_1$, \ or is regular outside them if $a=1$.

\section{Null limits}

The purpose of this paper is to investigate the limits of the above
solutions as $A\to\infty$. To maintain the inequality (\ref{rangemA}), it
is necessary to simultaneously scale the parameter $m$ to zero such that
$mA$ remains constant. The parameter $\varphi$ is unchanged by this
scaling. However, the parameters $Z_1$ and $Z_2$ become zero in these
limits, but their ratio remains a finite constant, and all the regularity
conditions are thus preserved.

Let us first consider the above null limit for the more familiar case
${\cal B}$. In this case, we find (expanding to terms in $Z_i^2$) that
the limits of (\ref{emuB}) are 
 \begin{equation}
\e^\mu_{\cal B}\to a, \quad {\rm everywhere}; \qquad
\e^\lambda_{\cal B}\to 
\left\{\matrix{a, & {\rm outside \ the \ null \ cone.} \cr  
\noalign{\medskip}
a\,{\displaystyle \bigg({Z_2\over Z_1}\bigg)^2}, 
& {\rm inside \ the \ null \ cone.} \cr} 
\right.  
 \end{equation} 
 Thus $\e^\mu_{\cal B}$ is constant everywhere, but there is a
discontinuity in $\e^\lambda_{\cal B}$ on the null cone \
$\rho^2+\zeta^2-\tau^2=0$. \ This corresponds to the presence of a
spherical impulsive gravitational wave exactly as described in a different
context in the previous paper \cite{PodGri00c}. Also, there is generally a
conical singularity on the axis of symmetry \ $\rho=0$. \ However, this
can be removed either inside or outside the null cone by an appropriate
choice of the constant $a$. For the choice \ \hbox{$a=Z_1/Z_2$}, \ the
axis is regular inside the null cone, but a conical singularity appears on
the axis outside it. This situation can be considered to describe the
``snapping'' of a cosmic string with deficit angle \ $(1-\beta)2\pi$, \
where 
 \begin{equation}
\beta= Z_2/Z_1 =\textstyle{{1\over2}
\,\big[\,1+\sqrt3\cot(\varphi+{1\over3}\pi)\,\big]} \,,
 \end{equation}
 so that
$\beta\in(0,1]$. \ Alternatively, for the choice \ $a=1$, \ the axis is
regular outside the null cone, but there is a conical singularity on the
axis inside. This describes a strut, or a conical singularity with excess
angle \ $(1-\beta^{-1})2\pi$, \ whose length is increasing in both
directions at the speed of light.

The null limit for the case ${\cal D}$ is, in fact, exactly equivalent to
that of the previous case ${\cal C}$. In this case, the null limits of
(\ref{emuD}) are 
 \begin{equation}
\e^\mu_{\cal D}\to a, \quad {\rm everywhere}; \qquad
\e^\lambda_{\cal D}\to 
\left\{\matrix{a, & {\rm outside \ the \ null \ cone.} \cr  
\noalign{\medskip}
a\,{\displaystyle \bigg({Z_1\over Z_2}\bigg)^2}, 
& {\rm inside \ the \ null \ cone.} \cr} 
\right.  
 \end{equation} 
 Again $\e^\mu_{\cal D}$ is constant everywhere, but there is a
discontinuity in $\e^\lambda_{\cal D}$ on the null cone corresponding
to the presence of a spherical impulsive gravitational wave. For the
choice \ \hbox{$a=Z_2/Z_1$}, \ the axis is regular inside the null cone,
but there is a conical singularity on the axis outside corresponding to a
snapped strut which has an excess angle \ $(1-\beta^{-1})2\pi$. \
Alternatively, for the choice \ $a=1$, \ the axis is regular outside the
null cone, but there is a conical singularity on the axis inside,
corresponding to an expanding cosmic string with deficit angle \
$(1-\beta)2\pi$.

In the above limit for the remaining case ${\cal A}$, the metric functions
become 
 \begin{equation}
\e^\mu_{\cal{A}} \to 
\left\{\matrix{a, \qquad \cr  
\noalign{\medskip}
{\displaystyle a\,{\rho^4\over(\zeta^2-\tau^2)^2}}, \cr} \right.
 \qquad
\e^\lambda_{\cal{A}} \to
\left\{\matrix{a, & \qquad {\rm outside \ the \ null \ cone.} \cr  
\noalign{\medskip}
{\displaystyle
b\,{\rho^4(\zeta^2-\tau^2)^2\over(\rho^2+\zeta^2-\tau^2)^8} },  &
\qquad {\rm inside \ the \ null \ cone.} \cr} \right.
 \end{equation}
 Thus, the region outside the null cone reduces to part of Minkowski
space, which is regular on the axis if $a=1$ (otherwise a conical
singularity remains). However, the equivalent limit for
$\e^\lambda_{\cal{A}}$ inside the null cone vanishes. In fact, a curvature
singularity appears on the null cone and the space-times in the two
regions cannot be connected. In order to obtain a finite limit for 
$\e^\lambda_{\cal{A}}$ inside the null cone, it is necessary to
rescale the parameter $a$ in this component as \
$a=b\,(4Z_1Z_2)^{-2}=9bA^8(4\cos^2\varphi-1)^{-2}$, \ where $b$ is held
constant. The resulting space-time in this region also contains a curvature
singularity on the axis of symmetry. This limit, and indeed the general
case ${\cal A}$, is not physically significant.

\section{Discussion}

The null limits of the cases ${\cal B}$ and ${\cal D}$ described in the
previous section are together identical to the equivalent null limit of
the Bonnor--Swaminarayan solution which is described in the previous paper
\cite{PodGri00c}. In that paper, we give the transformation of the metric
to appropriate continuous forms that are more appropriate for its physical
interpretation as a spherical impulsive gravitational wave generated by a
snapping or expanding cosmic string (with a deficit or excess angle). 
Using the transformations (22) and (24) in \cite{PodGri00c}, we obtain the
continuous metric in the Gleiser--Pullin form \cite{GlePul89}
 \begin{equation}
 \d s^2=4 \,\d {\cal U}\, \d {\cal V} 
 -\big({\cal V}-P\,{\cal U} \big)^2 \,\d \phi^2 
 -\big({\cal V}+P\,{\cal U} \big)^2 \,\d \psi^2 ,
 \label{GP}
 \end{equation} 
 where, for case ${\cal B}$
 \begin{equation}
 P= \left\{\matrix{ \Theta({\cal U}) +\beta^2\,\Theta(-{\cal U}) 
 &\qquad{\rm for} \quad a=\beta^{-1} & : \ {\rm snapping \ string} \cr
\noalign{\medskip}
 \beta^{-2}\,\Theta({\cal U}) +\Theta(-{\cal U}) 
&\qquad{\rm for} \quad a=1 & : \ {\rm expanding \ strut} \cr }
 \right.
 \end{equation}
 The equivalent metric for case ${\cal D}$ for a snapping strut or an
expanding string is obtained by replacing $\beta$ by $\beta^{-1}$ above.

It is also of interest to contrast the solution described here with that
of Aichelburg and Sexl \cite{AicSex71} in which a single Schwarzschild
black hole is boosted to the relativistic limit. In that case, a plane
impulsive gravitational wave is generated by a single null particle (the
Ricci tensor has a singular point on the wave surface). By contrast in
this case, the structure of the two black holes in the C-metric vanishes
in the null limit (the Ricci tensor vanishes everywhere on the null cone).
However, the strings remain, and the motion of their end points generates
impulsive spherical gravitational waves.

\section*{Acknowledgements}

This work was supported by a visiting fellowship from the Royal Society
and, in part, by the grant GACR-202/99/0261 of the Czech Republic.

\end{document}